\documentclass{aa}
\usepackage{graphicx}
\usepackage[varg]{txfonts}
\usepackage[pdfborder={0 0 0},colorlinks=true,citecolor=blue]{hyperref}

\begin{document}

\title{Probing the redshift evolution and sub-populations of binary neutron stars with the Einstein Telescope}
\titlerunning{Observing binary neutron stars with the Einstein Telescope}

\author{A. Toubiana \inst{1,2} \and I. Dvorkin \inst{3,4}}

\institute{
Dipartimento di Fisica “G. Occhialini”, Universit\'a degli Studi di Milano-Bicocca, Piazza della Scienza 3, 20126 Milano, Italy
\and
INFN, Sezione di Milano-Bicocca, Piazza della Scienza 3, 20126 Milano, Italy
\and
  Institut d'Astrophysique de Paris, UMR 7095, CNRS and Sorbonne Université, 98 bis boulevard Arago, 75014 Paris, France
\and
Institut Universitaire de France, Minist\`ere de l'Enseignement Sup\'erieur et de la Recherche, 1 rue Descartes, 75231 Paris Cedex F-05, France
  }

\abstract
{}
{The formation channels of binary neutron stars (BNSs) currently remain uncertain, but important information can be gathered by observing their mergers with gravitational-wave detectors. The processes that lead to BNS coalescence are encoded in the time-delay distribution between stellar binary formation and BNS coalescence, and therefore in the BNS merger rate. Moreover, the detection of GW190425 by LIGO/Virgo/KAGRA (LVK) suggests a sub-population of massive BNSs, possibly formed through unstable 'case BB' mass transfer with short merger delays. We investigate whether next-generation detectors such as the Einstein Telescope (ET) can constrain the time-delay distribution of BNSs and identify such sub-populations.}
{Using the latest LVK constraints, we generated mock ET catalogues that contain a mixture of light and heavy sub-populations. We modelled the redshift distribution of each sub-population as the convolution of the cosmic star formation rate with a time-delay distribution. We first considered a scenario where the time-delay distribution is common to all BNSs and follows a power law with indices $\alpha=-0.5,-1,-1.5$. In the second scenario, heavy BNSs have fixed short delays, while light BNSs follow power-law delays with the same set of indices. Hierarchical Bayesian analyses were then performed on catalogues of 100–5{,}000 events.
}
{With thousands of events, ET will be able to accurately characterise the time-delay distribution for the $\alpha=-0.5$ and $\alpha=-1$ cases. We find that with hundreds of detections from ET, we will be able to establish that the total mass distribution is bimodal. A few thousand events are sufficient to disentangle the redshift distributions of the two sub-populations for moderate time-delay indices ($\alpha_{\rm L}=-0.5$ or $-1$). For steeper indices ($\alpha_{\rm L}=-1.5$), the differences are more subtle and require larger catalogues, which was beyond what we could explore given our computational resources.}
{Next-generation detectors should enable the detection of multiple BNS sub-populations and their redshift evolution, and provide valuable insights into their formation pathways.}

   \keywords{neutron stars, binaries, gravitational waves }

\maketitle

\section{Introduction}

Direct observations of binary neutron stars (BNSs) come primarily from radio observations of binary pulsars in the Milky Way~\citep{2016ARA&A..54..401O}\footnote{In this paper, BNS refers to binary systems where both components are neutron stars.}, and thus probe the local population. The mass distribution of neutron stars (NSs) in these systems is found to be much narrower than that of the overall NS population~\citep{2018MNRAS.478.1377A,2016arXiv160501665A,2020RNAAS...4...65F,2020PhRvD.102f3006S}, and can be well described by a Gaussian distribution centred at $1.33 \
 M_{\odot}$ with a standard deviation of $0.09 \ M_{\odot}$~\citep{2012ApJ...757...55O,2016ARA&A..54..401O}, although there is some evidence of bimodality between recycled and non-recycled pulsars, the former having slightly larger masses~\citep{2019ApJ...876...18F}. This particularity of double radio pulsars could be attributed to the fact that the components of BNSs undergo much less accretion than neutron stars in other types of binaries, or to different masses at birth~\citep{2017ApJ...846..170T}.

Gravitational waves (GWs) provide a new means to directly observe these systems. The LIGO/Virgo/KAGRA (LVK) collaboration has so far confidently reported the detection of two BNSs: GW170817~\citep{2017PhRvL.119p1101A} and GW190425~\citep{2020ApJ...892L...3A}, at $z\sim 0.009$ and $z\sim 0.03$, respectively. Next-generation ground-based observatories such as the Einstein Telescope (ET)~\citep{2023JCAP...07..068B,2026JCAP...03..081A} and Cosmic Explorer (CE)~\citep{2021arXiv210909882E,2023arXiv230613745E}, expected to come online by the end of the next decade, should detect thousands to tens of thousands of BNSs out to redshift $z \sim 3$. The redshift distribution of BNS mergers encodes the convolution of the cosmic star formation rate (SFR) with the distribution of time delays between binary formation and coalescence. As such, it provides a direct probe of the physical processes that govern BNS formation. The orbital separation at the formation of the BNS, and therefore the time delay to coalescence, is determined by various stellar evolution processes, including mass transfer episodes, common envelope phases, and supernova explosions with associated natal kicks~\citep[e.g.][]{2018MNRAS.474.2937C,2018MNRAS.480.2011G,2018MNRAS.481.4009V,2019MNRAS.490.3740N,2021MNRAS.502.4877S}. Observations across a broad redshift range with next-generation detectors will therefore enable a reconstruction of the merger rate density and place strong constraints on the underlying time-delay distribution, offering a powerful means to discriminate between formation channels. These observations will also enable tests for the presence of multiple BNS sub-populations, as current data may already suggest. While the component masses of GW170817 are consistent with those of the Galactic population, GW190425 appears significantly heavier. This discrepancy is particularly evident in terms of the total mass: $3.4^{+0.03}_{-0.01}\,M_{\odot}$ for GW190425, compared to $2.66 \pm 0.12\,M_{\odot}$ for Galactic BNSs, which potentially points to the existence of a distinct, more massive sub-population.

This intriguing observation naturally raises the question of how GW190425 formed and why such heavy BNSs are not observed in the Milky Way. Dynamical formation in stellar clusters could, in principle, produce heavier BNSs if a massive neutron star exchanges its stellar companion with another neutron star; however, this channel is expected to contribute negligibly to the overall BNS merger rate~\citep{2020ApJ...888L..10Y,2018A&A...615A..91B}. Focusing on isolated binary evolution, some studies suggest that GW190425 could have formed through pathways similar to those of Galactic BNSs~\citep{2024A&A...691A.214Q,2025MNRAS.543..233N}, and that the spins and associated magnetic fields of heavy BNSs may render them effectively radio-invisible~\citep{2020ApJ...900...13S,2025ApJ...980..181C}. Alternatively,~\cite{2020MNRAS.496L..64R} proposed that massive binaries like GW190425 may originate from a distinct evolutionary channel characterised by much shorter delay times between the formation of the progenitor stars and BNS merger, potentially explaining their scarcity in the Galaxy. In this scenario, the system undergoes unstable `case BB' mass transfer from a helium star onto the first-formed neutron star~\citep{2002ApJ...572..407B,2003MNRAS.344..629D,2017ApJ...846..170T,2018MNRAS.481.4009V,2023MNRAS.524..426I}, triggering a second common-envelope phase between the neutron star and the remaining CO core. The resulting tighter orbital separation allows the binary to survive the second supernova kick and merge within a few million years. Following this idea,~\cite{2021ApJ...909L..19G} jointly fitted the radio and GW BNS populations and found mild evidence that GW190425 could belong to a fast-merging population comprising $8$–$79\%$ of all BNSs, with associated delay times between $5$ and $401\ {\rm Myr}$ (assumed to be the same for all heavy BNSs). We stress that other studies also find that such a discrepancy between the Milky Way and extragalactic populations can naturally arise in population synthesis, without the need for a distinct time-delay distribution~\citep{2020A&A...639A.123K}. As the exact pathway to produce GW190425 is still open to debate, in this work we explore the consequence of the scenario proposed by~\cite{2020MNRAS.496L..64R} for future observations. If this hypothesis is correct, we could expect to observe two sub-populations of BNSs with GWs: a light population and a heavy one, potentially exhibiting distinct redshift evolutions, which ET and CE could be able to probe.

Using the latest constraints from the LVK on the local BNS merger rate~\citep{2025arXiv250818083T}, we constructed mock catalogues of BNS observations on which we performed hierarchical Bayesian analyses to reconstruct the astrophysical distributions. First, considering the case of a single population, we illustrate the accuracy with which the time-delay distribution can be reconstructed. We then turn to the possibility of multiple sub-populations. We find that approximately $500$ events are sufficient to identify the presence of two sub-populations, while a few thousand detections allow us to distinguish their redshift distributions; this highlights the potential of next-generation facilities to constrain BNS formation channels.

\section{Binary neutron star population}\label{sec:pop}

Our population model builds upon the work of~\citet{2019ApJ...876...18F} and~\citet{2021ApJ...909L..19G}. We emphasise that it is not directly informed by binary stellar evolution calculations, but is designed to capture the phenomenology relevant to the case studied in this work. In Ref.~\citet{2021ApJ...909L..19G}, BNS systems are described as pairs composed of a recycled and a slowly spinning NS, each drawn from either a heavy (${\rm H}$) or a light (${\rm L}$) sub-population. Here, we further assumed that BNSs form preferentially as light and light or heavy and heavy, as appears to be the case for GW170817 and GW190425. Under this assumption, the individual NS masses, $m_1$ and $m_2$, associated with the heavy/light sub-populations are sampled from Gaussian distributions with means and standard deviations of $(\mu_{1,{\rm H}/{\rm L}}, \sigma_{1,{\rm H}/{\rm L}})$ and $(\mu_{2,{\rm H}/{\rm L}}, \sigma_{2,{\rm H}/{\rm L}})$, respectively. The numerical values of these parameters, adopted from~\citet{2021ApJ...909L..19G}, are listed in Table~\ref{tab:params}.

{
\renewcommand{\arraystretch}{1.3}
\begin{table}[hbtp!]
\caption{Parameters of our population models. }\label{tab:params}
\centering
   \begin{tabular}{c *{1}{c}}

   \cline{1-2}

  \multicolumn{1}{c}{Parameter}  &  \multicolumn{1}{|c}{Value} \\

    \cline{2-2}

    \hline

    \multicolumn{1}{c}{$\mu_{1,{\rm L}} $} & \multicolumn{1}{|c}{$1.34 \ M_{\odot}$} \\

    \multicolumn{1}{c}{$\sigma_{1,{\rm L}} $} & \multicolumn{1}{|c}{$0.02 \ M_{\odot}$} \\

     \multicolumn{1}{c}{$\mu_{2,{\rm L}} $  } & \multicolumn{1}{|c}{$1.29 \ M_{\odot}$} \\

     \multicolumn{1}{c}{$\sigma_{2,{\rm L}} $ } & \multicolumn{1}{|c}{$0.09 \ M_{\odot}$} \\

     \multicolumn{1}{c}{$\mu_{1,{\rm H}}$} & \multicolumn{1}{|c}{$1.47 \ M_{\odot}$} \\

    \multicolumn{1}{c}{$\sigma_{1,{\rm H}} $} & \multicolumn{1}{|c}{$0.15\ M_{\odot}$} \\

     \multicolumn{1}{c}{$\mu_{2,{\rm H}} $  } & \multicolumn{1}{|c}{$1.80 \ M_{\odot}$} \\

     \multicolumn{1}{c}{$\sigma_{2,{\rm H}}$ } & \multicolumn{1}{|c}{$0.15 \ M_{\odot}$} \\

     \hline

     \multicolumn{1}{c}{$\lambda$ } & \multicolumn{1}{|c}{$0.50$} \\

      \multicolumn{1}{c}{$\lambda_{\mathcal{R}}$ } & \multicolumn{1}{|c}{$0.69 ; \ 0.56; \ 0.49$} \\

     \hline

     \multicolumn{2}{c}{$\Lambda_{t,{\rm L}}$ } \\

     \hline

     \multicolumn{1}{c}{$\mu_{t,{\rm L}} $} & \multicolumn{1}{|c}{$2.63 \ M_{\odot}$} \\

    \multicolumn{1}{c}{$\sigma_{t,{\rm L}} $} & \multicolumn{1}{|c}{$0.092 \ M_{\odot}$} \\

    \hline

    \multicolumn{2}{c}{$\Lambda_{t,{\rm H}}$ } \\

     \hline

     \multicolumn{1}{c}{$\mu_{t,{\rm H}} $} & \multicolumn{1}{|c}{$3.27 \ M_{\odot}$} \\

    \multicolumn{1}{c}{$\sigma_{t,{\rm H}} $} & \multicolumn{1}{|c}{$0.21 \ M_{\odot}$} \\

    \hline

    \multicolumn{2}{c}{$\Lambda_{q,{\rm L}}$ } \\

     \hline

      \multicolumn{1}{c}{$\mu_{q,{\rm L}} $} & \multicolumn{1}{|c}{$1.00$} \\

    \multicolumn{1}{c}{$\sigma_{q,{\rm L}} $} & \multicolumn{1}{|c}{$0.08$} \\

    \hline

    \multicolumn{2}{c}{$\Lambda_{q,{\rm H}}$ } \\

     \hline

      \multicolumn{1}{c}{$\mu_{q,{\rm H}} $} & \multicolumn{1}{|c}{$0.82$} \\

    \multicolumn{1}{c}{$\sigma_{q,{\rm H}} $} & \multicolumn{1}{|c}{$0.20$} \\

    \hline

    \multicolumn{2}{c}{$\Lambda_{z,{\rm L}}$ } \\

     \hline

      \multicolumn{1}{c}{$\Delta t_{\rm min,L}$ } & \multicolumn{1}{|c}{$10 \ {\rm Myr}$} \\

      \multicolumn{1}{c}{$\Delta t_{\rm max,L}$ } & \multicolumn{1}{|c}{$13 \ {\rm Gyr}$} \\

      \multicolumn{1}{c}{$\alpha_{\rm L}$ } & \multicolumn{1}{|c}{$-0.5 ; \ -1 ; \ -1.5$} \\

      \hline

    \multicolumn{2}{c}{$\Lambda_{z,{\rm H}}$ } \\

     \hline

     \multicolumn{1}{c}{$\Delta t_{\rm H}$ } & \multicolumn{1}{|c}{$30 \ {\rm Myr}$} \\

   \end{tabular}
   \tablefoot{The first eight lines list the parameters of the population model of~\cite{2021ApJ...909L..19G}, which inspired our own. We then report the values of the hyperparameters used to generate the mock populations in our study, grouped by category. For completeness, we also provide the values of $\lambda_{\mathcal{R}}$, when considering two distinct sub-populations, that correspond to the three choices of $\alpha_{\rm L}$. When assuming a common redshift distribution, we use the same parameters as for the light sub-population. }
  \end{table}

  \begin{figure*}
\centering
\includegraphics[width=0.99\linewidth]{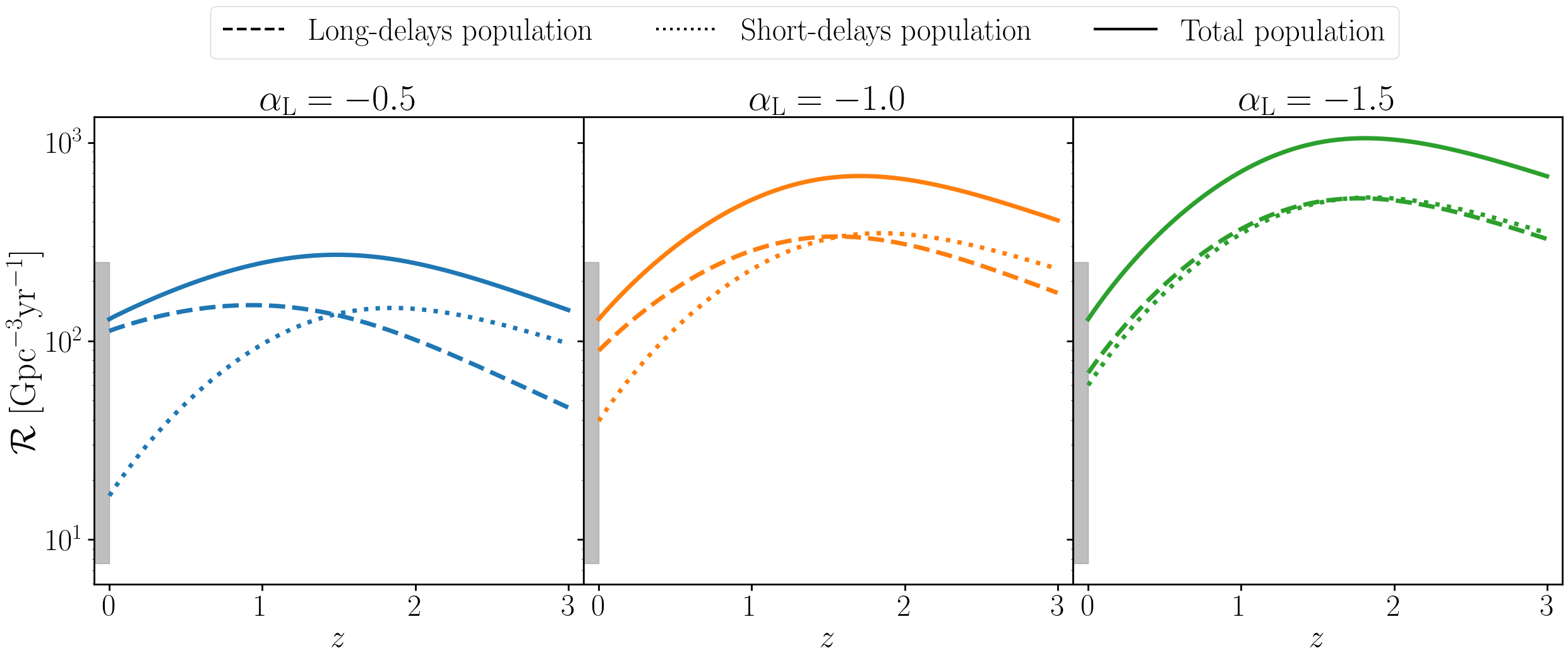}
   \caption{Merger rate of the long delays population (dashed lines), short delays population (dotted lines), and total population (full line) for three different hypotheses of the time-delays distribution. The rate is normalised at $z=0$ to the mean of the interval reported by the LVK following GWTC-4~\citep{2025arXiv250818083T}. The grey band at $z=0$ shows the $90\%$ credible interval reported by the LVK. In the case of a single population, the merger rate follows the long-delay curve, with the appropriate normalisation to match the total event rate.}\label{fig:rate_alpha}
 \end{figure*}

In this formulation, no explicit ordering between $m_1$ and $m_2$ is imposed. To prevent labelling ambiguities that may arise in the GW context, we reparametrised the model in terms of the total mass $m_t = m_1 + m_2$ and the mass ratio $q = \mathrm{min}(m_1, m_2) / \mathrm{max}(m_1, m_2) \leq 1$. The total mass is assumed to follow a normal distribution with a mean and standard deviation of
$\mu_{t,{\rm H}/{\rm L}} = \mu_{1,{\rm H}/{\rm L}} + \mu_{2,{\rm H}/{\rm L}}$ and
$\sigma_{t,{\rm H}/{\rm L}} = \sqrt{\sigma_{1,{\rm H}/{\rm L}}^2 + \sigma_{2,{\rm H}/{\rm L}}^2}$, respectively.
The mass ratio is modelled as a truncated log-normal distribution between 0 and 1, characterised by parameters $\mu_{q,{\rm H}/{\rm L}}$ and $\sigma_{q,{\rm H}/{\rm L}}$. The parameters of the distribution were chosen to qualitatively reproduce the population of \citet{2021ApJ...909L..19G}, as we show that it is indeed the case in Appendix~\ref{sec:comp_pop}. If we denote the fraction of light BNS with $\lambda$  and use \mbox{$\Lambda=(\lambda,\Lambda_{t,{\rm L}},\Lambda_{q,{\rm L}},\Lambda_{z,{\rm L}},\Lambda_{t,{\rm H}},\Lambda_{q,{\rm H}},\Lambda_{z,{\rm H}})$} to denote the full set of population parameters, i.e. the hyperparameters, our population model reads
\begin{align}
p(m_t, q, z | \Lambda) = & \lambda   \times p(m_t | \Lambda_{t,{\rm L}})  p(q | \Lambda_{q,{\rm L}})  p(z | \Lambda_{z,{\rm L}})  \nonumber \\
& + (1-\lambda) \times p(m_t | \Lambda_{t,{\rm H}})  p(q | \Lambda_{q,{\rm H}})  p(z | \Lambda_{z,{\rm H}}).\label{eq:pop_model}
\end{align}

The redshift distribution is given by
\begin{equation}
    p(z|\Lambda_{z,{\rm H}/{\rm L}}) \propto \frac{1}{1+z} \frac{{\rm d}V_c}{{\rm d}z} \mathcal{R}_{{\rm H}/{\rm L}}(z).\label{eq:p_of_z}
\end{equation}
Distributions were normalised between $z_{\rm min}=0$ and $z_{\rm max}=3$, as we fond that almost no BNS event is detected past redshift 3 with ET. In the above equation, $\frac{{\rm d}V_c}{{\rm d}z}$ is the differential comoving volume, and $\mathcal{R}_{{\rm H}/{\rm L}}(z)$ is the volumetric rate of events. The latter depends on the cosmic SFR, the BNS formation efficiency, $\eta_{{\rm H}/{\rm L}}$, and the time-delay distribution, $p_{{\rm H}/{\rm L}}(\Delta t)$:
\begin{equation}
\mathcal{R}_{{\rm H}/{\rm L}}(z) = \int \eta_{{\rm H}/{\rm L}}(t(z) + \Delta t) {\rm SFR}(t(z) + \Delta t)  p_{{\rm H}/{\rm L}}(\Delta t)  {\rm d}\Delta t.\label{eq:rate_mer}
\end{equation}

We considered two distinct scenarios for the time-delay distribution. First, we considered the case of a single BNS population, where we assumed a unique time-delay distribution that follows a power law with index $\alpha$, bounded between $\Delta t_{{\rm min}} = 10 \ {\rm Myr}$ and $\Delta t_{{\rm max}} = 13 \ {\rm Gyr}$. Although our mass model assumed a bimodal distribution to capture the high-mass end, our conclusions on the reconstruction of the time-delay distribution are largely insensitive to the precise form of the mass model. For instance, adopting a skewed distribution with an extended high-mass tail would yield comparable results, provided that the abundance of heavy systems is similar. For simplicity, we therefore adopted a single model for the mass distribution, namely the one described above.

In the second scenario, following~\citet{2020MNRAS.496L..64R} and~\citet{2021ApJ...909L..19G}, we assumed that the heavy sub-population merges on systematically shorter timescales than the light one. In this case, heavy binaries were assigned a fixed, short delay time of $\Delta t_{\rm H} = 30 \ {\rm Myr}$, modelled as $p_{\rm H}(\Delta t) = \delta(\Delta t - \Delta t_{\rm H})$, while light binaries were assumed to follow a power-law delay-time distribution with index $\alpha_{\rm L}$, bounded between $\Delta t_{{\rm L,min}} = 10 \ {\rm Myr}$ and $\Delta t_{{\rm L,max}} = 13 \ {\rm Gyr}$. In both cases we fond that values of the lower cutoff of the power-law time-delay distribution, $\Delta t_{{\rm L,min}}$, and/or the fixed delay of the heavy population, $\Delta t_{\rm H}$, below $\sim 100,{\rm Myr}$ have a negligible impact on the resulting redshift distributions. Hence, the results presented below are largely insensitive to the precise choice of these parameters.

We adopt the cosmic SFR from~\citet{2014ARA&A..52..415M}. The formation efficiency of each sub-population was assumed to be constant with redshift and thus acted as a normalisation factor. The overall normalisation was constrained by the total BNS merger rate inferred by the LVK Collaboration~\citep{2025arXiv250818083T}, while the relative contributions of each sub-population were set by the choice of the mixing fraction~$\lambda$. We emphasise that the mixing fraction entering the probability density function, $\lambda$, differs from the one entering the volumetric rate, $\lambda_{\mathcal{R}}$. If we define the latter such that
$\mathcal{R}(z) = \lambda_{\mathcal{R}} \mathcal{R}_{\rm L}(z) + (1 - \lambda_{\mathcal{R}})\mathcal{R}_{\rm H}(z)$,
the two are related by
\begin{equation}
\lambda_{\mathcal{R}} =
\frac{
\lambda \int_{z_{\rm min}}^{z_{\rm max}} \frac{\mathcal{R}_{\rm H}(z)}{1+z} \frac{{\rm d}V_c}{{\rm d}z}  {\rm d}z
}{
\int_{z_{\rm min}}^{z_{\rm max}}  \frac{ \lambda \mathcal{R}_{\rm L}(z) + (1-\lambda)\mathcal{R}_{\rm H}(z)}{1+z} \frac{{\rm d}V_c}{{\rm d}z}  {\rm d}z
}.
\end{equation}
Here, we assumed an equal contribution from both populations, i.e.~$\lambda=0.5$.
We summarise the parameters of our population in Table~\ref{tab:params}. For completeness, we also report there the corresponding values of $\lambda_{\mathcal{R}}$. Our values are compatible with the estimates of~\cite{2021ApJ...909L..19G} that $8$-$79\%$ of BNS at birth are rapidly merging (in terms of $\lambda_{\mathcal{R}}$).

 Since the time-delay distribution is poorly constrained, we explored different assumptions for the delay-time distribution of the slowly merging (light) population. Specifically, we fixed the minimum and maximum values of the time-delay range and vary the power-law index, both in the single population and distinct sub-population cases. Population synthesis models typically find $\alpha =-1$ \citep[e.g.][]{2018MNRAS.474.2937C}, although at specific progenitor metallicities the distribution can be much shallower \citep{2024MNRAS.535.2041D,2025A&A...693A.283P}, while observational-based estimates that use gamma-ray bursts or Galactic pulsars find much steeper values of $\alpha = -1.5$ to $-2$ for some sub-populations \citep{2022ApJ...940L..18Z,2025ApJ...982..179M}. In this work we considered $\alpha = -0.5, -1,$ and $-1.5$. The corresponding volumetric merger rates are shown in Fig.~\ref{fig:rate_alpha}.

In each case, the total rate is normalised to the mean of the $90\%$ credible interval on the local BNS merger-rate range reported by the LVK~\citep{2025arXiv250818083T}, between $7.6$ and $250 \ {\rm Gpc}^{-3}{\rm yr}^{-1}$, i.e. $129 \ {\rm Gpc}^{-3} {\rm yr}^{-1}$. Using the value of $1.16 \times 10^{-2} \ {\rm Mpc^{-3}}$ for the number density of Milky Way-equivalent galaxy (MWEG)~\citep{2008ApJ...675.1459K}, this rate corresponds to a merger rate per MWEG of $0.65$-$ 22\ {\rm MWEG}^{-1}{\rm Myr}^-1$. The grey band at $z=0$ shows the credible interval reported by the LVK.
The short-delay population is rescaled by a different normalisation factor for each model, since for a fixed $\lambda$ the corresponding volumetric-rate mixing fraction $\lambda_{\mathcal{R}}$ depends on $\alpha_{\rm L}$. Larger values of $\alpha_{\rm L}$ assign greater weight to long delays, thereby enhancing the contrast with the short-delay population.

  \begin{figure}
\centering
\includegraphics[width=0.99\linewidth]{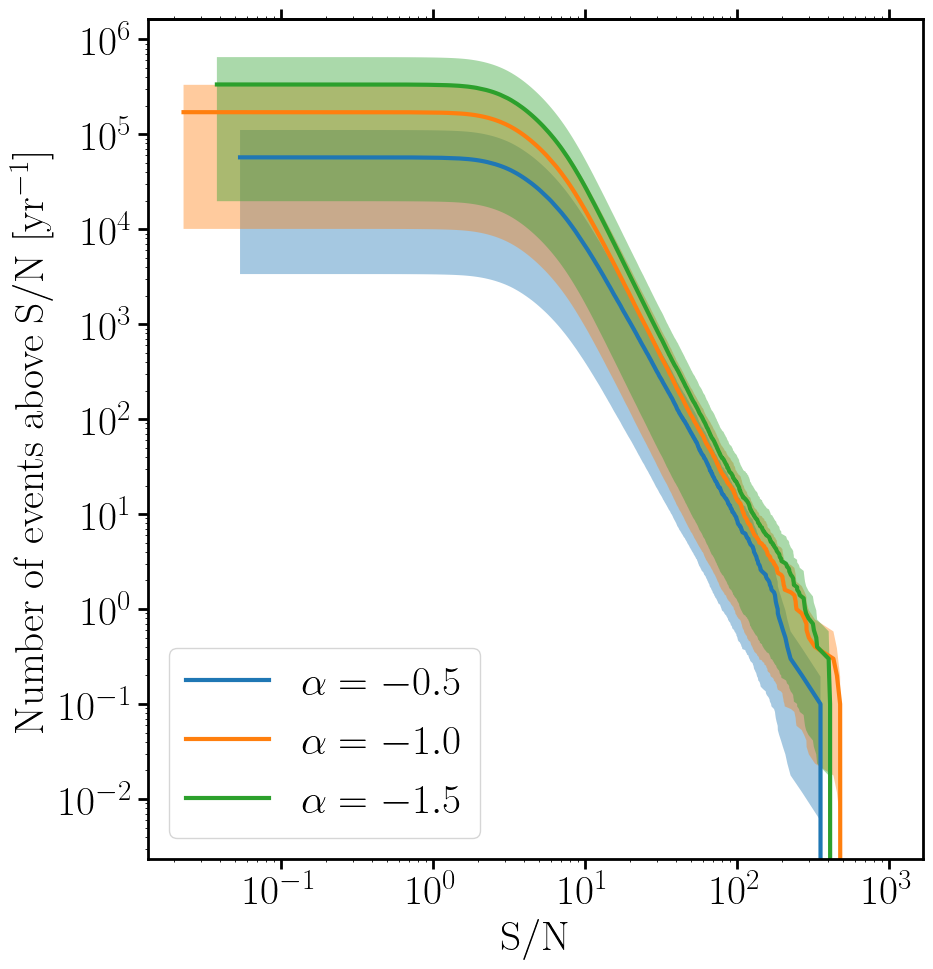}
   \caption{Number of events above a given S/N  threshold per year in the single population scenario for the three values of $\alpha$ considered in this work. }\label{fig:snrs_alpha}
 \end{figure}

\section{Mock catalogues and data analysis}

We considered a triangular configuration for ET with a $10 \ {\rm km}$ arm length and adopt the \texttt{ET-10} (10 km-xylophone) sensitivity curve~\citep{DanilishinZhang:2023_ET-0304B-22} implemented in \texttt{GWBENCH}~\citep{2021CQGra..38q5014B}.
Our mock detection and parameter estimation pipeline was based on the approach outlined in~\cite{2020ApJ...891L..31F} and \citet{2023ApJ...955..107F}. Differences arising in our implementation are discussed in Appendix~\ref{app:pe}.

Figure~\ref{fig:snrs_alpha} shows the cumulative number of detections as a function of signal-to-noise ratio (S/N) threshold for the three values of $\alpha$ in the single population case. The shaded regions represent the uncertainty associated with the local BNS merger rate reported by the LVK~\citep{2025arXiv250818083T}, corresponding to the lower and upper bounds of the $90\%$ credible interval.
If we assume a S/N threshold of eight in the single population scenario, we estimate that ET will detect between approximately $680$–$22{,}310$, $1{,}710$–$56{,}280$, and $3{,}090$–$101{,}630$ BNS mergers per year for $\alpha = -0.5$, $-1$, and $-1.5$, respectively. In the case of two sub-populations, we estimate these numbers to be $1{,}180$–$38{,}920$, $2{,}370$–$77{,}950$, and $3{,}320$–$109{,}150$ BNS mergers per year for $\alpha_{\rm L} = -0.5$, $-1$, and $-1.5$, respectively. In all cases, $\sim 93\%$ of detected BNS are within redshift 2. These estimates are compatible with~\cite{2023JCAP...07..068B} after accounting for the fact that they assumed a local merger rate of $250 \ {\rm Gpc}^{-3}{\rm yr}^{-1}$, which lies at the higher end of our range.

Based on these projections, we generated mock catalogues containing $100$, $500$, $1{,}000$, and $5{,}000$ detected events, with ten realisations for each case. We then performed hierarchical Bayesian inference to recover the population hyperparameters. These catalogue sizes are consistent with the expected ET detection rate, towards its lower end, while remaining computationally tractable. Current computational resources, in terms of both processing time and memory requirements, prevent us from extending the analysis to the tens of thousands of detections that ET is expected to observe. For this reason, we restricted our study to catalogues of up to $5{,}000$ events. The hierarchical Bayesian framework is described in detail in Appendix~\ref{app:hba}; here, we summarise the population modelling choices relevant to this work.

Our population prior is given by a mixture model as in Eq.~\eqref{eq:pop_model}. The component distributions in total mass and mass ratio are modelled as Gaussian and truncated log-normal distributions, respectively. We adopted flat priors on the mixture fraction, as well as on the means and standard deviations of these distributions. For the redshift dependence, we explore two complementary approaches.

We assumed that the redshift evolution of each sub-population follows Eqs.~\eqref{eq:p_of_z} and~\eqref{eq:rate_mer}.  In the inference, we modelled the delay-time distribution of both sub-populations as a power law, i.e. we do not assume any of them to have a fixed delay time. The delta-function limit is recovered when $\Delta t_{\rm max} \rightarrow \Delta t_{\rm min}$.
The priors adopted are:
\begin{itemize}
    \item $\Delta t_{\rm min}$: log-flat between $1\,{\rm Myr}$ and $1\,{\rm Gyr}$;
    \item $\Delta t_{\rm max}-\Delta t_{\rm min}$: log-flat between
    $10^{-3} \ {\rm Myr}$ and $13.5 \ {\rm Gyr}$\footnote{The limit $\Delta t_{\rm max} = \Delta t_{\rm min}$ is not included in our prior; however, intervals shorter than $1\,{\rm Myr}$ are in practice hardly distinguishable from a delta function.};
    \item $\alpha$: flat between $-3$ and $0$.
\end{itemize}

When investigating the ability of ET to identify distinct BNS sub-populations, after performing the hierarchical Bayesian inference, we computed the probability that the mass distribution is bimodal, $p_{\rm bimodal}$, as follows. For each hyperposterior sample, we drww $2{,}000$ $(m_t,q)$ samples and apply Hartigan's dip test~\citep{Hartigan1985DipTest} using the implementation of~\cite{diptest}. This test quantifies whether the empirical probability distribution function of the samples exhibits a `dip,' which would indicate the presence of two distinct modes. The test returns a $p$ value that can be interpreted as the probability that the distribution is unimodal. Repeating this for all hyperposterior samples yields a distribution of $p$ values for each hierarchical Bayesian analysis.
We then computed $p_{\rm bimodal}$ as the fraction of $p$ values below $0.05$, which provides an estimate of the probability that the mass distribution is bimodal with more than $95\%$ confidence. We verified that the number of $(m_t,q)$ samples drawn per hyperparameter sample was sufficient, in the sense that increasing it does not affect the results.
Since Hartigan's dip test is defined for one-dimensional data, we rotated the $(m_t,q)$ samples for each hyperparameter sample using a rotation matrix, and retained the rotation that yields the lowest $p$ value (i.e. the highest probability of bimodality). This procedure allowed us to identify which linear combination of $m_t$ and $q$ exhibits the strongest bimodality. In practice, we find that the angle that maximises the bimodality was approximately zero, which indicates that the total mass distribution itself is the main source of bimodality.

The probability that the two redshift distributions are different, $p_{\rm dist}$, was estimated as follows. We randomly drew pairs of hyperparameters from the posterior samples and computed the Kolmogorov–Smirnov (KS) statistic between the resulting redshift distributions of the two sub-populations. This 'between-population' KS distribution captures the typical differences between the two sets of distributions. To construct a background distribution, we also computed two 'within-population' KS distributions by comparing different samples from the same sub-population, which provides a reference for variations expected from measurement uncertainty (detector noise and a finite number of events). We then determined the $5\%$ quantile of the between-population KS distribution and find its corresponding quantiles in the within-population KS distributions. The maximum of these two quantiles is defined as $p_{\rm dist}$. It measures whether the difference between the sub-populations exceeds the typical measurement-induced variations in at least one subpopulation, indicating that the distributions can be reliably distinguished.

\begin{figure}
\centering
\includegraphics[width=0.99\linewidth]{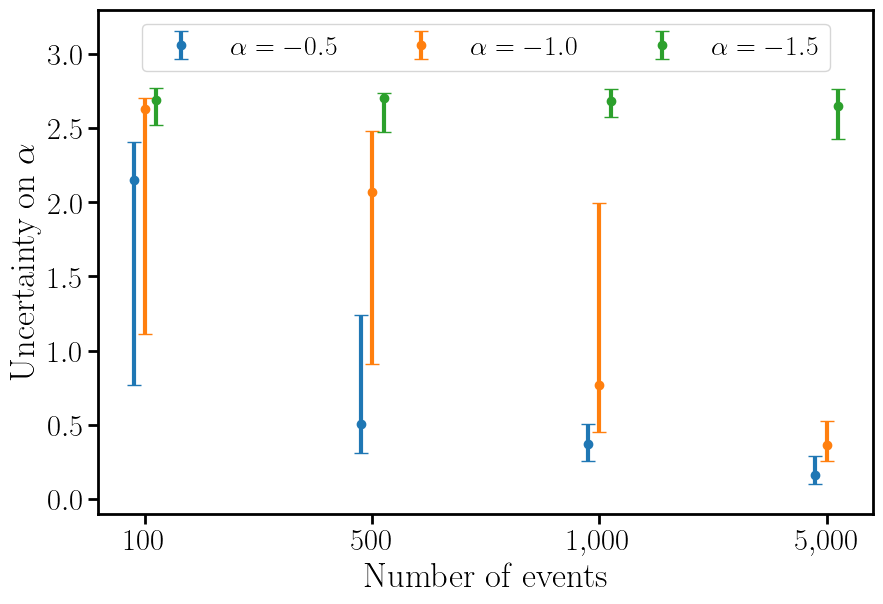}
   \caption{Credible intervals on the width of the $90\%$ credible interval for the power-law exponent, as a function of the number of events and for different values of $\alpha$.}\label{fig:delta_alpha}
 \end{figure}

 \begin{figure}
\centering
\includegraphics[width=0.99\linewidth]{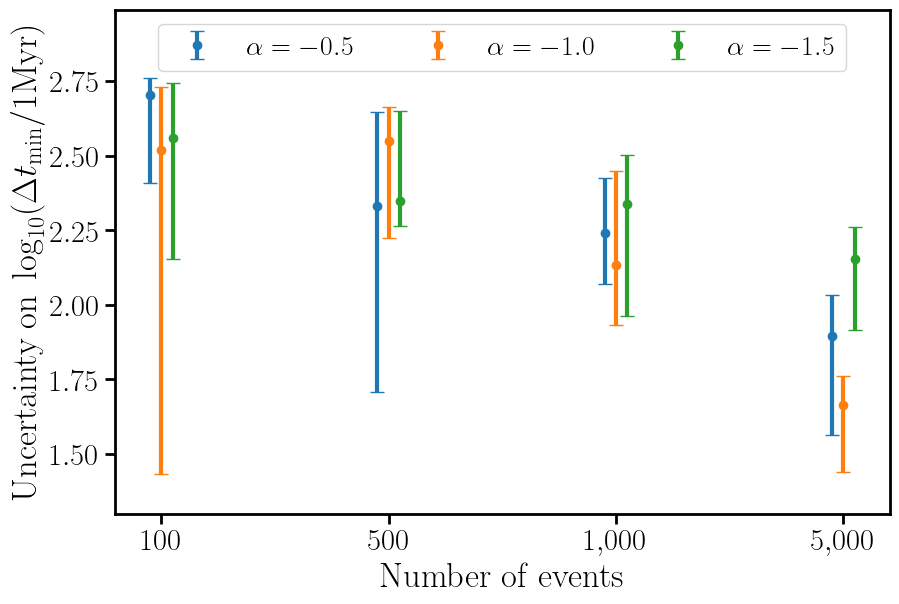}
   \caption{ Credible intervals on the width of the $90\%$ credible interval for the minimum time delay, as a function of the number of events and for different values of $\alpha$.}\label{fig:delta_tmax}
 \end{figure}

 \begin{figure}
\centering
\includegraphics[width=0.99\linewidth]{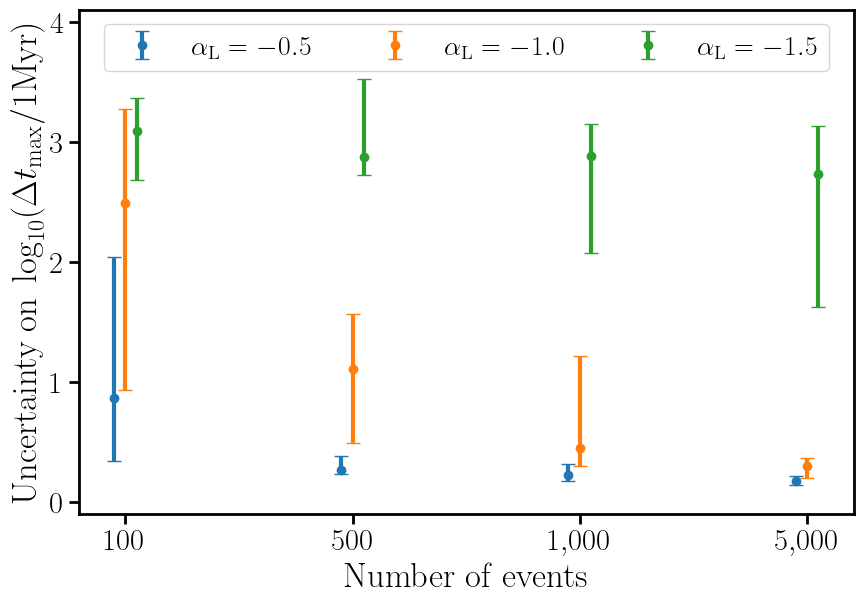}
   \caption{Credible intervals on the width of the $90\%$ credible interval for the maximum time delay, as a function of the number of events and for different values of $\alpha$.}\label{fig:delta_tmin}
 \end{figure}

\section{Results}

 \begin{figure}
\centering
\includegraphics[width=0.99\linewidth]{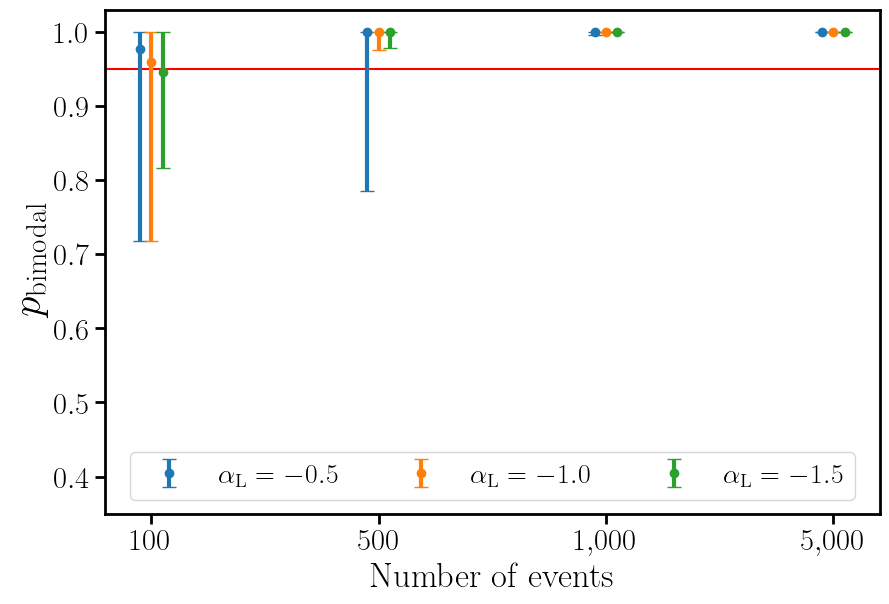}
   \caption{Credible intervals on the confidence that the mass distribution is bimodal as a function of the number of events and for the different values of $\alpha_{\rm L}$. The horizontal red line corresponds to a probability of 0.95.   }\label{fig:p_multi_alpha}
 \end{figure}

 \begin{figure*}
\centering
\includegraphics[width=0.99\linewidth]{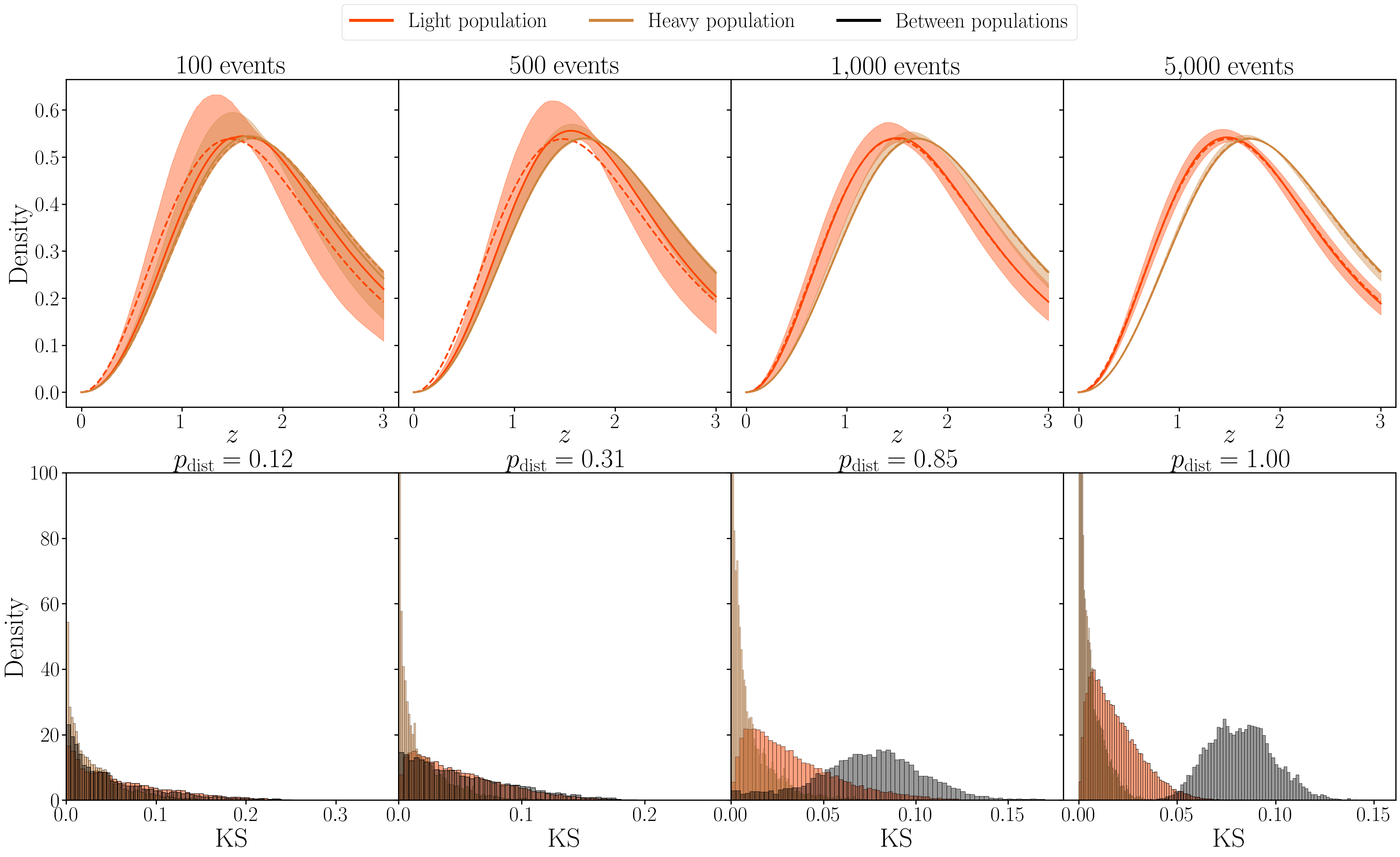}
   \caption{Upper panel: Reconstructions of the redshift distributions for increasing catalogue sizes. Coloured bands show the $90\%$ credible intervals for the light (red) and heavy (golden) populations. Solid lines indicate the median reconstructions, while dashed lines mark the true distributions. For the heavy population, the dashed and full lines superimpose almost perfectly. Lower panel: Distribution of within-population and between-population KS statistics. The value on top shows the probability that the two heavy and light population have different redshift distributions. }\label{fig:z_strong}
 \end{figure*}

 \begin{figure}
\centering
\includegraphics[width=0.99\linewidth]{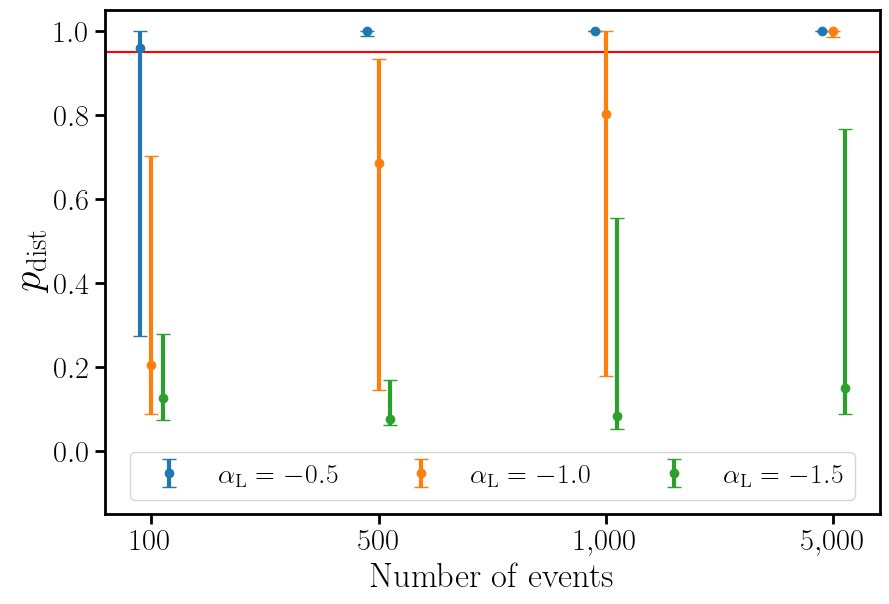}
   \caption{Credible intervals on the confidence with which we can determine that the redshift distribution of the light and of the heavy population are different as a function of the number of events and for the different values of $\alpha_{\rm L}$. The horizontal red line corresponds to a probability of 0.95.   }\label{fig:p_dist_alpha}
 \end{figure}

\subsection{Single population}

We first discuss the results in the case of a common redshift distribution, focusing on how well the parameters of the time-delay distribution can be constrained. Figure~\ref{fig:delta_alpha} shows the uncertainty on the power-law index, $\alpha$, as a function of the number of events. We find that, with $1{,}000$ observations, $\alpha$ can already be constrained within $\sim 0.5$ for an injected value of $\alpha = -0.5$, while a similar accuracy is achieved with $\sim 5{,}000$ events for $\alpha = -1.0$.

For $\alpha = -1.5$, however, the power-law index remains essentially unconstrained even after $5{,}000$ events. In this case, the redshift distribution closely resembles that obtained for a fixed delay, as shown in the right panel of Fig.~\ref{fig:p_multi_alpha}. This behaviour is reflected in the large uncertainty on the maximum time delay, shown in Fig.~\ref{fig:delta_tmax}, which remains uncertain at the level of several gigayears even after $5{,}000$ observations.

In contrast, for the other scenarios, the maximum delay time is very well constrained, within a few megayears, as it strongly affects the number of sources at $z \sim 0$, where the observational constraints are tightest. The minimum time delay is constrained to within a few hundred megayears in all cases, which is consistent with our observation in Sect.~\ref{sec:pop} that varying $\Delta t_{\rm min}$ by a few megayears has little impact on the results.

\subsection{Sub-populations}

We start by discussing the recovery of the mass distribution. With 100 events, we can already confidently measure that $\lambda \neq 0$ and $\lambda \neq 1$, which indicates that the population cannot be described by a single component. In Appendix~\ref{app:mass_rec} we show examples of how the total mass and mass ratio distributions are recovered as the number of events increases. Figure~\ref{fig:p_multi_alpha} displays the 90\% confidence interval of $p_{\rm bimodal}$ (centred on the median) as a function of the number of events. With $500$ detections, we can already establish in most cases that the total mass distribution is bimodal, and with $1{,}000$ detections this can be determined unambiguously, regardless of $\alpha_{\rm L}$.

Next, we turn to the distinction between the two redshift distributions. The upper panel of Fig.~\ref{fig:z_strong} shows representative examples of the reconstructed redshift distributions for $\alpha_{\rm L}=-1$ as the number of events increases. The lower panel shows the distribution of KS statistics, both between-population and within-population. The value above the bottom panel indicates $p_{\rm dist}$. This figure provides a visual interpretation of our method for estimating the probability that the two redshift distributions differ: as the confidence bands overlap less, the between-population KS distribution is moved towards higher values than the within-population ones, and the computed $p_{\rm dist}$ increases.
We observe that the redshift distribution of the heavy population is typically better constrained than that of the light population, as a consequence of their larger S/N.

Our results for $p_{\rm dist}$ are shown in Fig.~\ref{fig:p_dist_alpha}. For $\alpha_{\rm L}=-0.5$, we can already establish that the two redshift distributions differ with more than $95\%$ confidence using $500$ events. For $\alpha_{\rm L}=-1$, this level of significance is reached in over $95\%$ of realisations with $5{,}000$ events. For $\alpha_{\rm L}=-1.5$, the distinction becomes more challenging, and it appears that many more events are needed in order to distinguish the distributions, in agreement with the results found in the single population case.

\section{Conclusions}
Despite major progress over the last decade, the formation pathways of BNSs remain uncertain. In particular, the number of mass transfer episodes and the criterion for their stability have a strong influence on the properties of the resulting population yet are still not fully understood. The redshift distribution of BNS mergers provides an observational handle on BNS formation processes, since it encodes the distribution of time delays between binary formation and coalescence and therefore the initial BNS orbital separation. Moreover, BNS masses could also be correlated with the formation channel. The detection of GW190425 by LVK highlighted the possibility that a significant fraction of BNSs may form through unstable 'case BB' mass transfer, resulting in much shorter delay times than the rest of the population~\citep{2020MNRAS.496L..64R,2021ApJ...909L..19G}. In this scenario, two BNS sub-populations with distinct redshift evolutions could coexist. In this paper, we explore the ability of future ground-based facilities such as the ET to characterise the time-delay distribution of BNSs and to probe the existence of distinct sub-populations.

Using the latest estimate of the BNS merger rate by the LVK~\citep{2025arXiv250818083T}, we generated mock BNS populations consisting of an equal mixture of light and heavy sub-populations, inspired by the parametric models of~\cite{2019ApJ...876...18F} and~\citet{2021ApJ...909L..19G}. Assuming that the redshift distribution of each sub-population is given by the convolution of the SFR with a time-delay distribution, we considered two scenarios. In the first, both light and heavy BNSs share a common time-delay distribution described by a power law, with index $\alpha_{\rm L} = -0.5, -1$, and $-1.5$. In the second, heavy BNSs are assigned a fixed, short time delay, while light BNSs follow a power-law distribution with the same set of indices. Independently of this choice, we find that ET should observe at least a few thousand BNS mergers over its lifetime, with estimated detection rates ranging from $1{,}710$ to $56{,}280$ events per year in the fiducial $\alpha_{\rm L}=-1$ case for a common redshift distribution.

We then considered multiple realisations of catalogues ranging from 100 up to $5{,}000$ events and performed hierarchical Bayesian analyses to reconstruct the astrophysical distributions.
Crucially, when considering the possibility of having two distinct sub-populations, we did not constrain either sub-population to a fixed delay time in the inference, allowing for a more general and flexible model. We find that for $\alpha = -0.5$ and $-1$, the power-law index can be constrained within $\sim 0.5$ after $1{,}000$ and $5{,}000$ observations, respectively. The minimum delay time is constrained within a few hundred megayears, while the maximum delay time can be determined to within a few megayears. For $\alpha = -1.5$, however, the parameters remain poorly constrained, as the resulting redshift distribution closely resembles that obtained for a fixed delay of $\sim 100\ {\rm Myr}$.

In the case of two sub-populations, for all $\alpha_L$ values, we find that with $500$ events we can identify the bimodality of the BNS total mass distribution in most cases, and with $1{,}000$ events we can do so in practically all cases. Moreover, a few thousand events are sufficient to disentangle the redshift distributions of the two sub-populations if $\alpha_{\rm L}=-0.5$ or $-1$. For $\alpha_{\rm L}=-1.5$, the difference between the two distributions is very small, and more observations would be required to distinguish them, if at all. We could not explore this further due to limited computational resources. This highlights a crucial point: current population analysis frameworks will face significant challenges when the number of observations reaches several thousand.}

Our study shows that next-generation ground-based detectors will provide information on the processes leading to BNS coalescence, and could reveal a sub-population of heavy BNSs merging on much shorter timescales than the Galactic population. The most likely scenario behind it would be unstable `case BB' mass transfer, which tightens the binary orbit, allowing the system to remain bound after the second supernova and resulting in shorter merger delays. Detecting such a population would provide insight into BNS formation channels, accessible only through GW observations, since rapidly merging heavy BNSs would be largely absent in the Milky Way, explaining the lack of Galactic binaries as massive as GW190425. We emphasise once again that, although this scenario provides a viable explanation for the discrepancy between GW190425 and the Galactic population, it is not the only one capable of doing so. Alternatively, if heavy BNSs exist in our Galaxy but are radio-invisible~\citep{2020ApJ...900...13S,2025ApJ...980..181C}, the Laser Interferometer Space Antenna (LISA) should enable them to be found  in the Milky Way~\citep{2021MNRAS.502.5576K}.

An important caveat of our analysis is that it was limited to two sub-populations. For instance, allowing for a light and heavy coupling would introduce a third sub-population, although the (currently limited) data tend to favour mostly heavy and heavy and light and light coupling. Moreover, several formation pathways may contribute to the BNS population~\citep{2017ApJ...846..170T,2018MNRAS.481.4009V}, potentially resulting in a more complex distribution. Neglecting this could severely bias population reconstruction when using parametric or astrophysical models for the population prior~\citep{2021ApJ...910..152Z,2021PhRvD.104h3027T,2023ApJ...955..127C}. Future work will explore more comprehensive analyses that combine population synthesis models~\citep{2025A&A...693A.283P} with multi-dimensional non-parametric approaches, such as that of~\cite{2025ApJ...994L..52T}.

\begin{acknowledgements}

We are thankful to S. Borhanian, T. Bruel and M. Quartin for fruitful discussions. A.T. is supported by MUR Young Researchers Grant No. SOE2024-0000125, ERC Starting Grant No.~945155--GWmining, Cariplo Foundation Grant No.~2021-0555, MUR PRIN Grant No.~2022-Z9X4XS, Italian-French University (UIF/UFI) Grant No.~2025-C3-386, MUR Grant ``Progetto Dipartimenti di Eccellenza 2023-2027'' (BiCoQ), and the ICSC National Research Centre funded by NextGenerationEU.

\end{acknowledgements}

 \bibliographystyle{aa}
\bibliography{Ref}

\begin{appendix}

\section{Comparison between population parametrisations}\label{sec:comp_pop}

\begin{figure}[h!]
\centering
\includegraphics[width=0.8\linewidth]{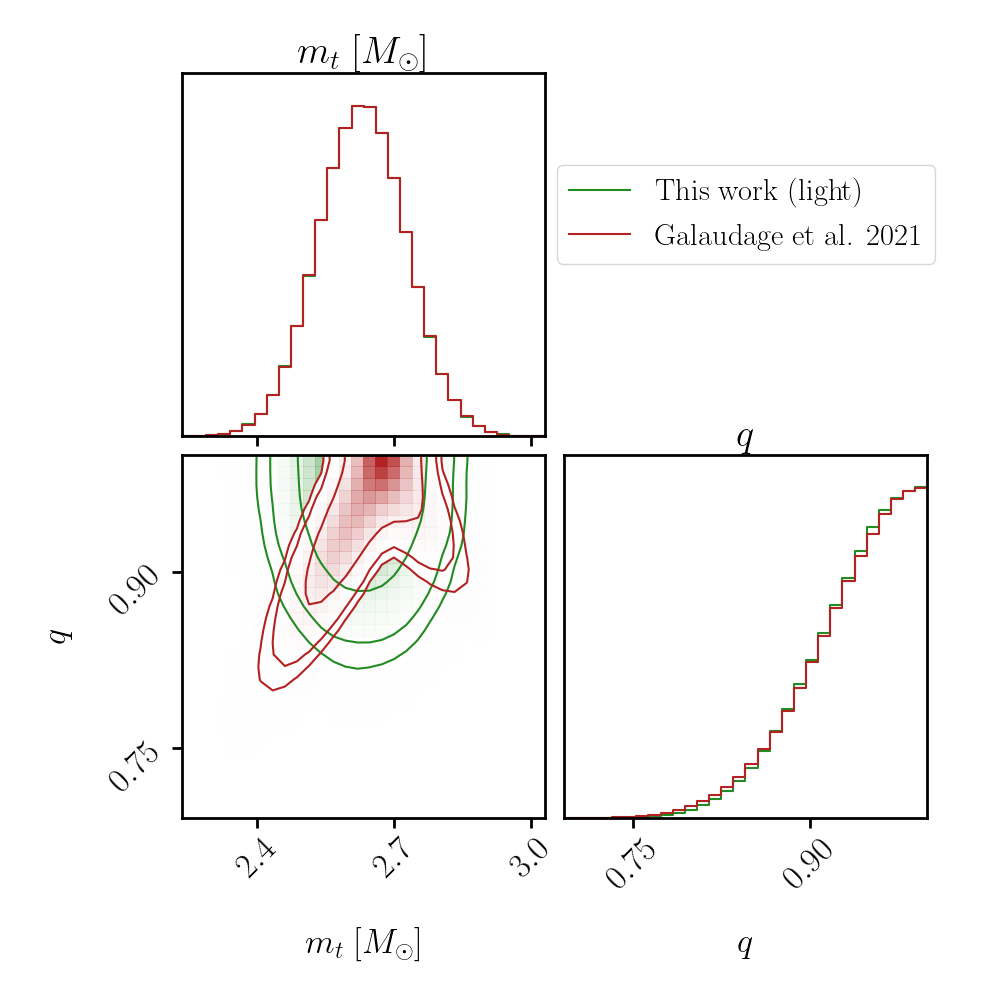}
   \caption{Comparison between the parametrisation of~\cite{2021ApJ...909L..19G} for the light population and ours in terms of total mass and mass ratio.    }\label{fig:comp_pop_light}
 \end{figure}

\begin{figure}[h!]
\centering
\includegraphics[width=0.8\linewidth]{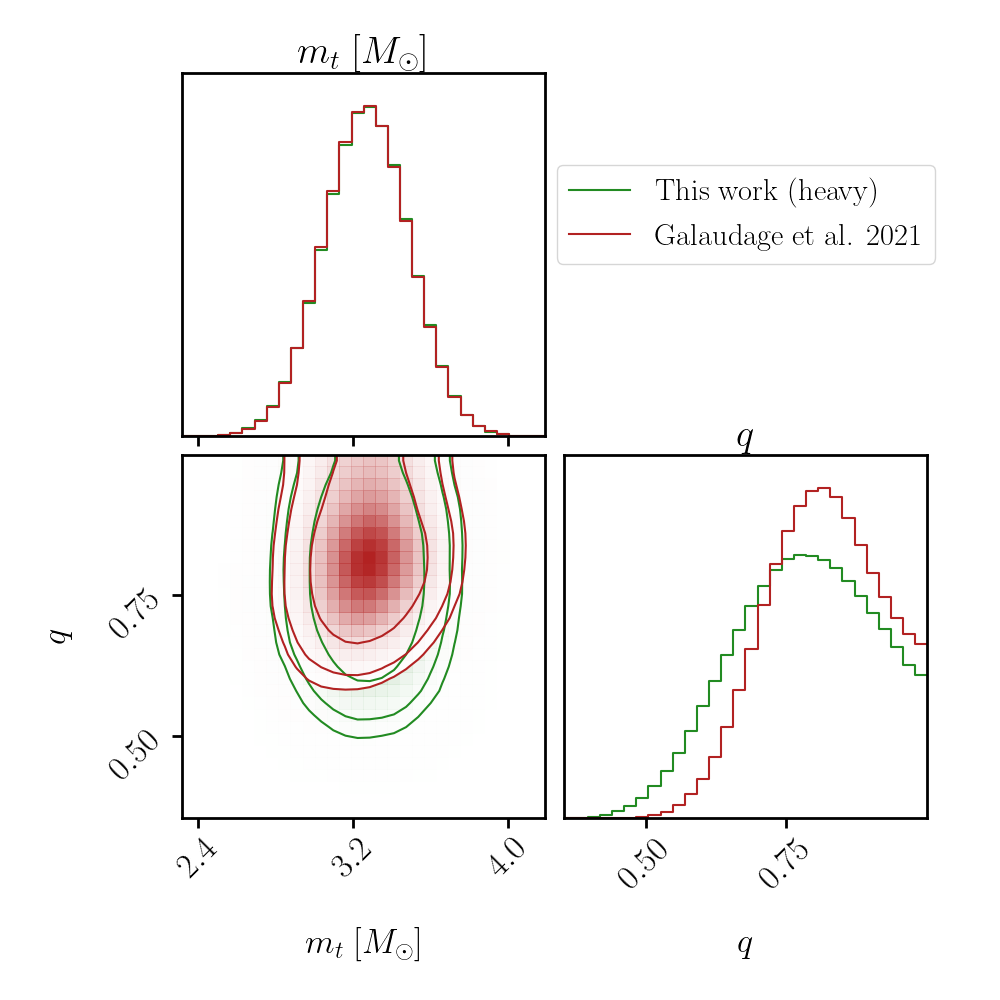}
   \caption{Comparison between the parametrisation of~\cite{2021ApJ...909L..19G} for the heavy population of ours in terms of total mass and mass ratio.  }\label{fig:comp_pop_heavy}
 \end{figure}

Figures~\ref{fig:comp_pop_light} and~\ref{fig:comp_pop_heavy} compare our population model with that of~\cite{2021ApJ...909L..19G}, with the extra condition that only light/light and heavy/heavy pairings are allowed. The total mass distributions are identical, as both models describe them as the sum of two Gaussians. The corresponding mass-ratio distributions are qualitatively similar. The discontinuity visible in the $(m_t, q)$ correlation for the light population originates from the fact that in~\cite{2021ApJ...909L..19G}, the component masses are not ordered; enforcing $q \leq 1$ thus introduces this artificial feature. Still, our model has a similar support to the original distribution.

\section{Mock detection and parameter estimation} \label{app:pe}

In their appendices,~\cite{2020ApJ...891L..31F} and~\cite{2023ApJ...955..107F} outline a prescription for generating mock gravitational-wave (GW) catalogues that self-consistently incorporate parameter estimation uncertainty and selection effects, while also reproducing the expected scaling of measurement errors with the inverse of the S/N. Their approach leverages the fact that the S/N, $\rho$, can be written as
\begin{equation}
\rho = \frac{F(\theta ')}{D_L},\label{eq:snr}
\end{equation}
where $\theta'$ denotes all binary parameters entering the S/N computation except for the luminosity distance $D_L$ (detector frame masses, spins, sky location...). For fixed $\theta'$, this expression allows one to transform straightforwardly between $D_L$ and $\rho$.

The key idea is to treat the S/N itself as a sampling parameter during mock parameter estimation, rather than the luminosity distance. This makes it possible to enforce that the measured S/N, $\tilde{\rho}$, is normally distributed around the true S/N, with unit variance: $\tilde{\rho} \sim \mathcal{N}(\rho,1) $, which reflects the expected behaviour of a matched-filter S/N in the case of a single detector~\citep{1993PhRvD..47.2198F}.
Selection effects are then included as follows: for each source, the true S/N is computed and scattered according to the above distribution to obtain $\tilde{\rho}$; the event is retained only if $\tilde{\rho}$ exceeds a chosen S/N threshold, $\rho_{\rm th}$.

For retained events, posterior samples of the S/N, ${\hat{\rho}}$, are then drawn from a normal distribution centred at $\tilde{\rho}$ with unit variance.
To model measurement uncertainty on the remaining parameters $\theta '$, ``noisy'' values $\tilde{\theta} '$ are generated by sampling from a normal distribution with variance proportional to $\tilde{\rho}^{-1}$, mimicking the typical scaling of parameter errors with inverse S/N. Posterior samples ${\hat{\theta} '}$ are obtained by further drawing from normal distributions centred at $\tilde{\theta} ' $ with variances also proportional to $\tilde{\rho}^{-1}$. The $\{\hat{\theta}',\hat{\rho}\}$ samples are then transformed back into  $\{\hat{\theta}',\hat{D_L}\}$ samples via Eq.~\eqref{eq:snr}. The proportionality constants for the measurement errors on $\theta '$ were calibrated against mock parameter estimation runs.

The problem can be simplified by noting that the S/N may be expressed as
\begin{equation}
\rho = \rho_{\rm opt}  \Theta,
\end{equation}
where $\rho_{\rm opt}$ is the optimal S/N for a source located directly overhead the detector with face-on inclination, and $\Theta$ encodes the dependence on sky location, polarisation, and inclination. For a single detector, $\Theta$ is bounded by $0\leq \Theta \leq 1$, while for ET the range becomes $0 \leq \Theta_{\rm ET} \leq \sqrt{3}$. Denoting by $F_{+,i}$ and $F_{\times,i}$ the antenna pattern functions of the three equivalent ET detectors, as given in~\cite{2012PhRvD..86l2001R}, the factor $\Theta_{\rm ET}$ can be written as
\begin{equation}
\Theta_{\rm ET} = \sqrt{\sum_{i=1}^{3} \left( F_{+,i}^2 + F_{\times,i}^2 \right)}.
\end{equation}
This expression generalises the single-detector relation derived in~\cite{1993PhRvD..47.2198F}, but is strictly valid only for the case of a triangular ET configuration, where all three detectors are co-located.
We note that in a network of multiple detectors, the distribution of the measured S/N, $\tilde{\rho}$, deviates slightly from the simple normal form assumed in the single-detector case~\citep{2023PhRvD.108d3011E,2024CQGra..41l5002G}. However, for the purposes of this study, we neglect this correction and adopt the single-detector approximation for simplicity.

Our approach differs from that of~\cite{2020ApJ...891L..31F,2023ApJ...955..107F} in two main respects. First, although we enforce hard prior boundaries when drawing the posterior samples ${\hat{\theta}'}$, we do not impose such constraints when generating the ``noisy'' parameter values $\tilde{\theta}'$. This choice reflects the fact that noise can make it appear as the likelihood peaks outside of the prior limits, which manifests as railing of the posterior against the prior boundary. This is also simpler, as it means that the likelihood is truly a Gaussian and not a truncated Gaussian, avoiding the need to adjust the sampling procedure to account for a modified functional form.

Second,~\cite{2020ApJ...891L..31F,2023ApJ...955..107F} adopt $\theta' = (\mathcal{M}_{c,d}, \eta, \Theta)$ as sampling parameters, where $\mathcal{M}_{c,d}$ is the detector-frame chirp mass and $\eta$ is the symmetric mass ratio. However, we find that sampling in the symmetric mass ratio leads to large variances in the Monte Carlo estimators used when evaluating the population likelihood (see App.~\ref{app:mass_rec}). This issue arises because our underlying population model favours nearly equal-mass systems, while sampling in $\eta$ tends to underpopulate the region near $q \sim 1$, as also discussed in~\cite{Farah2022_GWMockCat}. To mitigate this, we sample directly in the mass ratio $q$ instead of $\eta$.
We assume that uncertainties in the inferred mass ratio scale as $0.15\rho_{\rm th}/\rho$, which is a conservative estimate based on the measurement uncertainties reported for GW170817~\citep{2017PhRvL.119p1101A} and GW190425~\citep{2020ApJ...892L...3A}, particularly when tight spin priors are used. Although the mapping from detector-frame chirp mass to total mass is less problematic, we likewise choose to sample directly in the detector-frame total mass, adopting a similarly conservative uncertainty scaling of $0.1\rho_{\rm th}/\rho$. For $\Theta$ we use the same as~\cite{2020ApJ...891L..31F} and~\citet{2023ApJ...955..107F}: $0.21\rho_{\rm th}/\rho$.

We compute S/Ns using the \texttt{GWBENCH} package~\citep{2021CQGra..38q5014B}, setting the spins to 0 and using the IMRPhenomXHM waveform~\citep{2020PhRvD.102f4002G}. For the S/N threshold, we take $\rho_{\rm th}=8$.

\section{Hierarchical Bayesian analysis}\label{app:hba}

We denote with $\theta$ the set of parameters describing a GW event, $p(d|\theta)$ the single event likelihood and $p(\theta|\Lambda)$ the population prior on $\theta$, which depends on hyperparameters $\Lambda$ that we wish to infer.
The population likelihood for observing $N_{\rm obs}$ events \mbox{$\{d\} = \{d_1,\dots,d_{N_{\rm obs}}\}$}, marginalised over the event rate is~\cite{2019MNRAS.486.1086M,2022hgwa.bookE..45V}
\begin{equation}\label{eq:std_no_rate}
p(\{d\}|\Lambda) = \prod_i^{N_{\rm obs}}\int   \frac{p(d_i|\theta)p(\theta|\Lambda)}{p_{\rm det}(\Lambda)}\  {\rm d}\theta\,.
\end{equation}
We have introduced the selection function defined via
\begin{align}
    p_{\rm det}(\Lambda) =  \int p_{\rm det}(\theta) p(\theta|\Lambda) \ {\rm d}\theta , \\
   p_{\rm det}(\theta) = \int_{d > {\rm threshold}} p(d|\theta) \ {\rm d} d,
\end{align}
where the second integral is restricted to realisations $d$ that exceed the detection threshold (defined via the chosen ranking statistic), here the measured matched filter S/N $\hat{\rho}$.

Alternatively, if we do not wish to marginalise over the rate, we can write
\begin{equation}\label{eq:std_no_rate}
p(\{d\}|\Lambda) = e^{- N(\Lambda) p_{\rm det}(\Lambda)}\prod_i^{N_{\rm obs}}\int \ p(d_i|\theta)\frac{{\rm d}N}{{\rm d\theta}}(\Lambda) {\rm d}\theta  \,.
\end{equation}
where $\frac{{\rm d}N}{{\rm d\theta}}$ is the differential number of events, $N(\Lambda)=\int  \frac{{\rm d}N}{{\rm d\theta}}(\Lambda) \ {\rm d}\theta$ is the total expected number of events, and we have the relation $\frac{{\rm d}N}{{\rm d\theta}}= N(\Lambda)p(\theta|\Lambda)$.

In practice, the single-event likelihood can be written in terms of the posterior and the parameter estimation prior via Bayes’ theorem. The integral over $\theta$ is evaluated using Monte Carlo integration, based on posterior samples obtained as described in App.~\ref{app:pe}.

Similarly, the selection function is estimated via an injection campaign in which events are drawn from an injection distribution $\pi_{\rm inj}(\theta)$, and only those satisfying the detection criterion are retained, forming the set ${\theta_{\rm det}}$. The detection probability is then computed using importance sampling:
\begin{equation}
p_{\rm det}(\Lambda) = \sum_{\theta_{\rm det}} \frac{p(\theta_{\rm det} \mid \Lambda)}{\pi_{\rm inj}(\theta_{\rm det})}.
\end{equation}

As discussed in~\cite{2023MNRAS.526.3495T}, the finite number of samples used to evaluate the integrals in population analyses can introduce significant variance in the Monte Carlo estimators, which may in turn lead to biases. To mitigate this issue in the estimation of the selection function, we choose the injection prior to match the true population, thereby reducing the variance in the importance sampling weights. Additionally, in line with the recommendations of~\cite{2023MNRAS.526.3495T}, we impose a threshold on the total variance of the log-likelihood and discard hyperparameters exceeding this limit.

The posterior on $\Lambda$ is then obtained through Bayes' theorem: $p(\Lambda|\{d\})\propto p(\{d\}|\Lambda)\pi(\Lambda)$, with the priors described in Sec.~\ref{sec:pop} and in the next one. The sampling is performed with the \texttt{Eryn} sampler\footnote{Publicly available at \url{https://github.com/mikekatz04/Eryn}.}~\citep{2023MNRAS.526.4814K}.

\section{Reconstruction of the mass distribution}\label{app:mass_rec}

\begin{figure*}[!]
\centering
\includegraphics[width=0.99\linewidth]{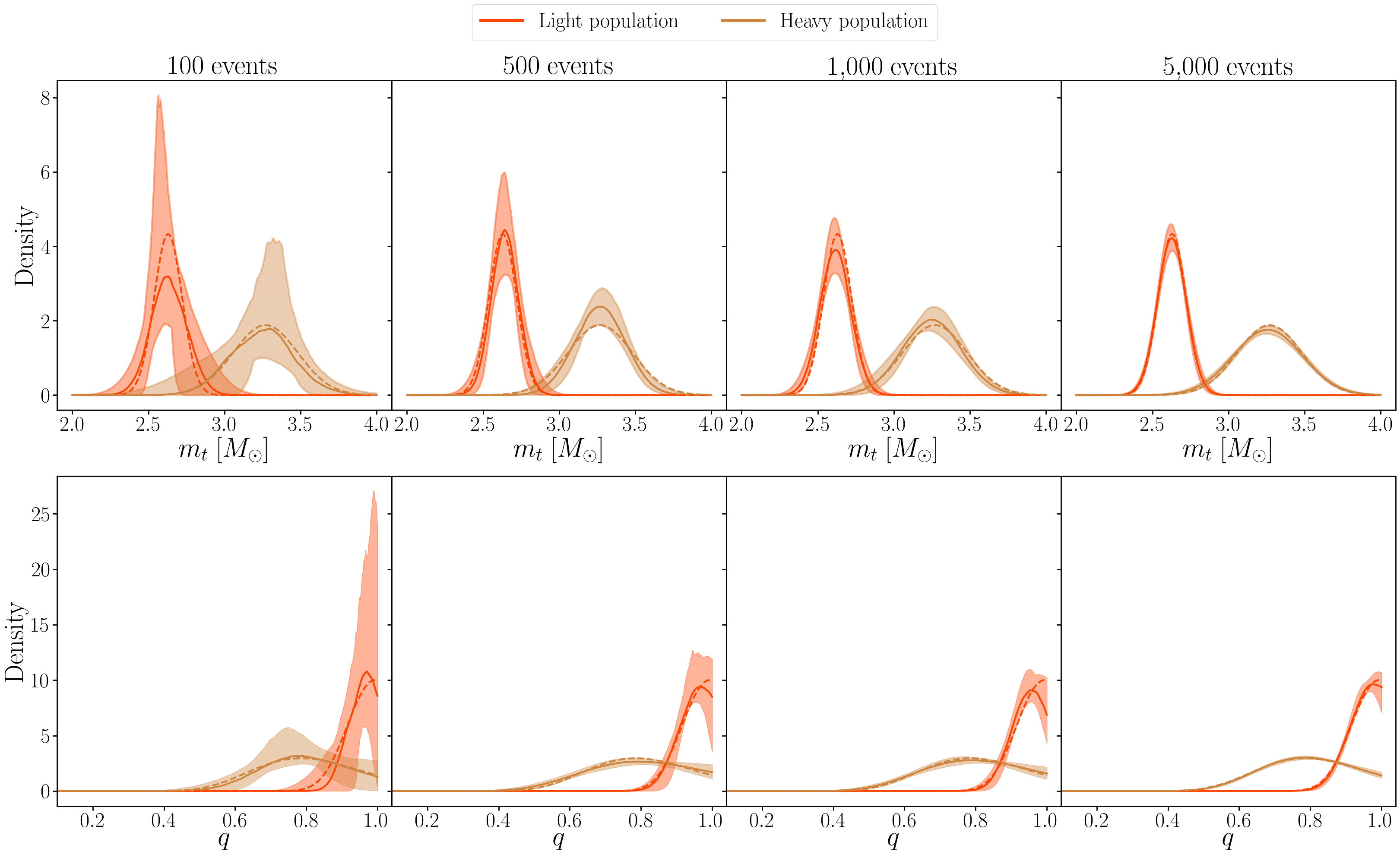}
   \caption{Reconstructions of the total mass and mass ratio distributions for increasing catalogue sizes.
Coloured bands show the $90\%$ credible intervals for the light (red) and heavy (golden) populations. Solid lines indicate the median reconstructions, while dashed lines mark the true distributions.}\label{fig:mtot_q}
 \end{figure*}

Figure~\ref{fig:mtot_q} shows representative examples of how the measurement of the total mass and the mass ratio distribution improves as the number of events increases. These plots correspond to realisations of the $\alpha_{\rm L}=-1$ case; however, the choice of $\alpha_{\rm L}$ has little impact on this result.

\end{appendix}

\end{document}